\documentclass{ws-fnl}
\usepackage{cite}
\usepackage{graphicx}
\begin{document}

\markboth{Authors' Names}{ (Paper's Title)}

\catchline{}{}{}{}{}

\title{Information fractal dimension of mass function}

\author{Chenhui Qiang}

\address{Institute of Fundamental and Frontier Science, University of Electronic Science and Technology of China, 610054 Chengdu, China \\ Yingcai Honors College, University of Electronic Science and Technology of China, Chengdu, 610054, China\\
2019270101007@std.uestc.edu.cn}

\author{Yong Deng\footnote{Corresponding author}}

\address{Institute of Fundamental and Frontier Science, University of Electronic Science and Technology of China, 610054 Chengdu, China \\
School of Education, Shaanxi Normal University, Xi'an, 710062, China \\
School of Knowledge Science, Japan Advanced Institute of Science and Technology, Nomi, Ishikawa 923-1211, Japan \\
Department of Management, Technology, and Economics, ETH Zürich, Zurich, Switzerland \\
dengentropy@uestc.edu.cn}

\author{Kang Hao Cheong\footnote{Corresponding author}}

\address{Science, Mathematics and Technology Cluster, Singapore, University of Technology and Design (SUTD), S487372, Singapore \\
SUTD-Massachusetts Institute of Technology International Design Centre, Singapore \\
kanghao\_cheong@sutd.edu.sg}

\maketitle

\begin{history}
\received{(received date)}
\revised{(revised date)}
\end{history}
\begin{abstract}
Fractal plays an important role in nonlinear science. The most important parameter to model fractal is fractal dimension. Existing information dimension can calculate the dimension of probability distribution. However, given a mass function which is the generalization of probability distribution, how to determine its fractal dimension is still an open problem of immense interest. The main contribution of this work is to propose an information fractal dimension of mass function. Numerical examples are illustrated to show the effectiveness of our proposed dimension. We discover an important property in that the dimension of mass function with the maximum Deng entropy is $\frac{ln3}{ln2}\approx 1.585$, which is the well-known fractal dimension of Sierpiński triangle.

\end{abstract}

\section{Introduction}
Fractal is very common in nature\ \cite{Mandelbrot1998The,jin2019general}, first proposed to measure the length of the coast. As the field progresses, it is now widely used across many different fields, such as time fractal\ \cite{wang2019remark,abu2019application}, the calculation and geometry\ \cite{he2018fractal,stanojevic2020fuzzy,arqub2019application}, and chaotic systems\ \cite{slimane2017designing,lin2017parameter}. Irrefutably, the fractal dimension plays an vital role in fractal theory. To date, a lot of measures \cite{wen2021review} has been proposed to determine the fractal dimension, including Hausdorff dimension\ \cite{lopes2009fractal}, information dimension\  \cite{zhu2020power,bian2018identifying}, correlation dimension\ \cite{lacasa2013correlation}, and multi-fractal dimension \ \cite{wei2016multifractality,wen2020vital}.

The uncertainty measurement is a topic of immense interest because it can be applied to the uncertain environment. Many algorithms have been proposed, including probability theory\ \cite{jaynes2003probability}, Dempster-Shafer evidence theory\  \cite{dempster2008upper,dempster1967upper,shafer1976mathematical} and belief structure\ \cite{yager2016maxitive,yager2018fuzzy}. The information volume of a probability distribution can be measured by Shannon entropy\ \cite{shannon1948mathematical}. The mass function in evidence theory is a generalization of probability set to describe the uncertainty environment which is also called basic probability assignment (BPA). The uncertainty of mass function can be measured by Deng entropy\ \cite{deng2016deng,Deng2020ScienceChina}.
Compared with probability distribution, mass function has been widely applied due to its ability in dealing with uncertain information. Many measures and parameters are developed about mass function, such as entropy\ \cite{Xiao2019EFMCDM,Xue2021Interval}, negation\ \cite{Xiao2020maximum}, correlation coefficient\ \cite{jiang2018Correlation} and information quality\ \cite{li2021multisource}. The information volume of the mass function has been studied recently\ \cite{Deng2020InformationVolume,deng2021fuzzymembershipfunction}. 

However, how to determine the fractal dimension of mass function is still an open problem. In this paper, we have proposed information dimension of mass function based on Deng entropy and fractal theory, which can be further applied in decision-making\ \cite{Liao2020}.
Importantly, we discover that the dimension of mass function with the maximum Deng entropy is 1.585, which is the same as the fractal dimension of Sierpinski triangle.

Entropy is very important in complex systems\ \cite{phillips2006maximum,ZhangINS2021,pincus1991approximate}. There are many different kinds of entropy function, such as Tsallis entropy\ \cite{tsallis1998role,Tsallis2021} and Renyi entropy\ \cite{van2014renyi}. In information theory, Shannon entropy plays an important role\ \cite{wang2017rumor,babajanyan2020energy}. As part of our literature review, we will introduce these concepts briefly. 
\subsection{Shannon entropy}

Given a probability distribution $P=\left \{p_{1},\ p_{2},...,\ p_{n}\right \}$, Shannon entropy is defined as follows\ \cite{1948A}
\begin{equation}
     H_{S}=-\sum_{i}p_{i}log(p_{i}).
\end{equation}
If and only if $p_{i}=\frac{1}{n}$, Shannon entropy reaches maximum
\begin{equation}
    H_{maxS}=log(n). 
\end{equation}

\subsection{Fractal and information dimension}

Fractal has been widely studied\ \cite{djennadi2021fractional} and fractal dimension is also attract a lot of focues\ \cite{yilmaz2020multiscale,ziaukas2017fractal}. For example, the fractal dimension of Sierpinski triangle is $\frac{ln3}{ln2}$ \ \cite{Ettestad2018THE}, the fractal dimension of Koch curve is $\frac{ln4}{ln3}$ \ \cite{2007Bounds}, and the fractal dimension of Cantor sets is $\frac{ln2}{ln3}$\ \cite{2003Hausdorff}. Besides, information dimension, as a kind of fractal dimension, plays a critical role in dealing with probability distribution. Below is a brief introduction of information dimension.

The information dimension is defined as follows$\ $\cite{peitgen2006chaos}
\begin{equation}
    D=\lim_{\varepsilon\rightarrow 0}\frac{\sum_{i=1}^{N} P_{i}(\varepsilon )lnP_{i}(\varepsilon )}{ln(\varepsilon) },
\end{equation}
where the numerator is Shannon entropy, $\varepsilon$ is the side length of the measured box, $P_{i}(\varepsilon)$ is the probability of the measured object falling into the $i$th box. $N$ represents $N$ measured boxes. 

\subsection{Mass function}

Mass function is an important aspect of evidence theory, which is an extension of probability theory.
$\Theta$ denotes the framework of discernment in evidence theory\ \cite{dempster1967upper,shafer1976mathematical},
\begin{equation}
\Theta=\{\omega_1,\cdots,\ \omega_N\},
\end{equation}
the power set of $\Theta$ is $2^{\Theta}$,
\begin{equation}
\begin{split}
2^{\Theta}=&\{A_{1},\ A_{2},...,\ A_{2^{N}}\} \\
=&\{\emptyset,\ \{\omega_1\},\cdots,\ \{\omega_N\},\ \{\omega_1,...,\ \omega_i\},\cdots,\ \{\Theta\}\},
\end{split}
\end{equation}


 The mass function is defined as follows\ \cite{dempster1967upper,shafer1976mathematical}
\begin{equation}
    m:2^{\Theta}\rightarrow [0,1].
\end{equation}
This mapping satisfies
\begin{equation}
    m(\emptyset)=0,
\end{equation}
\begin{equation}
   \quad \sum_{A\in 2^{\Theta}}m(A)=1.
\end{equation}

where $A_{i}$ is called focal element when $m(A_{i})>0$.
Mass function can be seen as a generalization of probability distribution, which is more efficient in dealing with uncertainty\ \cite{deng2021fuzzymembershipfunction,Xiao2020GIQ}.

\subsection{Deng entropy and maximum Deng entropy}

Recently, a new entropy called Deng entropy has been proposed to measure uncertainty of mass function. 
Given a mass function: $(\left \{A_{1},...,\ A_{2^{N}}\right \}:\left \{m(A_{i}),\ i=1,...,2^{N}\right \})$, its Deng entropy is obtained as
\begin{equation}
    H_{D}=-\sum_{A \in 2^{\Theta}}m(A)log(\frac{m(A)}{2^{|A|}-1}),
\end{equation}
where $|A|$ is the cardinal of focal element $A$.

When mass function is Bayesian structure\ \cite{shafer1976mathematical}, Deng entropy degenerates to Shannon entropy.

When the mass function satisfies the condition
\begin{equation}
    m(A_{i})=\frac{2^{|A_{i}|}-1}{\sum _{i}2^{|A_{i}|}-1},
\end{equation}
where $m(A_{i})$ is the mass function for $A_{i}$ and $i=1,2,...,2^{N}-1$, Deng entropy reaches the maximum, which is shown as follows\ \cite{Deng2020ScienceChina}
\begin{equation}
    H_{maxD}=-\sum_{i}m(A_{i})log(\frac{m(A_{i})}{2^{|A_{i}|}-1})=log\sum_{i}({2^{|A_{i}|}-1}).
\end{equation}

In probability theory, Shannon entropy reaches maximum when all results have equal probability. However, in evidence theory, Deng entropy postulates that when the uncertainty of mass function reaches the maximum, multiple subsets should have more assignment. 

The rest of the paper is organized as follows. Section 2 presents information dimension of mass function. Some numerical examples are given in Section 3 to illustrate the effectiveness of our proposed dimension. Finally, Section 4 concludes the paper.

\section{A new proposal for information dimension of mass function}

In this section, the new dimension is being proposed and we begin with some fundamental definitions. For convenience of the reader, we also provide the proof to the properties that are being discussed.

\begin{definition}
For a framework of discernment $
\Theta$, the power set is $2^{\Theta}=\{A_{1},\ A_{2},...,\ A_{2^{N}}\}$, a mass function is $m(A)$. Its information dimension is defined as follows,
 \begin{equation}
    D_{m}=\frac{H_{D}}{log\sum_{i}(2^{|A_{i}|}-1)^{m(A_{i})}},
\end{equation}
where $H_{D}$ is Deng entropy, $(2^{|A_{i}|}-1)$ is the size of power set of the focal element $A_{i}$.

\noindent \textbf {Property 1}: When $m(A_{i})=1,\  |A_{i}|=1$, both 0 in numerator and denominator of Equation (12), we defined $D_{m}$ as follows,

\begin{equation}
     D_{m}=0\  (m(A_{i})=1,\ |A_{i}|=1).
\end{equation}

\noindent \textbf {Proof}:

Given a framework of discernment $\Theta=\{\omega_1,\cdots,\ \omega_N\}$, a mass function is $m(A_{i})=a,\ m(A_{1})=...=m(A_{i-1})=m(A_{i+1})=...=m(A_{2^{N}-2})=b$, where $0<a<1,\ 0<b<1,\ a+(2^{N}-1)*b=1,\ |A_{i}|=1$.

When $a\rightarrow 1,\ b\rightarrow 0$, the mass function degenerates into $m(A_{i})=1,\ |A_{i}|=1$. According to Equation (9),
\begin{equation}
    \lim_{a\rightarrow 1}-m(A_{i})log\frac{m(A_{i})}{2^{|
A_{i}|}-1}=\lim_{a\rightarrow 1}-a\cdot log(a)\rightarrow 0
\end{equation}

\begin{equation}
    \lim_{b\rightarrow 0^{+}}-m(A_{j})log\frac{m(A_{j})}{2^{|
A_{j}|}-1}=\lim_{b\rightarrow 0^{+}}-b\cdot log(\frac{b}{2^{|
A_{j}|}-1}) \rightarrow  0
\end{equation}

\begin{equation}
\begin{split}
      \lim_{a\rightarrow 1, b\rightarrow 0}H_{D}=&-\sum_{A \in 2^{\Theta}}m(A)log(\frac{m(A)}{2^{|A|}-1})\\
      =&(-a\cdot log(a)+\sum_{j}-b\cdot log(\frac{b}{2^{|
A_{j}|}-1})) \rightarrow 0
\end{split}
\end{equation}
 where $1\leq j\leq 2^{N}-1\ and\ j\neq i$, according to Equation (12), we have
\begin{equation}
    D_{m}\rightarrow 0
\end{equation}


The special case with the condition $(m(A_{i})=1, \ |A_{i}|=1)$ in Property 1 indicates that the information is deterministic, so its information dimension is 0.
  
\noindent \textbf {Property 2}: When mass function degenerates into probability distribution: $(\left \{A_{1},...,\ A_{N}\right\})$: $(\left \{P_{i} >0,\ i=1,...,N\right \})$, where $|A_{i}|=1$. Equation (12) can be rewritten as follows.

\begin{equation}
   D_{m}=\frac{H_{D}}{log\sum(2^{|A_{i}|}-1)^{m(A_{i})}}=\frac{\sum_{i}-P_{i}log(P_{i})}{log\sum(2^{1}-1)^{P_{i}}}=\frac{H_{S}}{log(N)}=D_{p},
\end{equation}
where $H_{S}$ is Shannon entropy. For denominator, due to the elementary event is no longer split in probability theory, which is represented by the value of any exponent of 1 is 1. In other words, the split of each singleton is itself. 
\end{definition}

In the case of one mass function or probability distribution, only a number is obtained by Equation (12). However, in the next section when the mass function changes in a certain regularity, the number either stays the same or converges gradually to a constant, indicating that there is a scale invariance between Deng entropy and the splitting of mass function. Therefore, the number is used to represent this property and named as information fractal dimension.

\section{Numerical examples}

In this section, we provide some numerical examples to better illustrate the definition of the proposed dimension. In order to verify the results easily, all examples below use $log2$ for the purpose of calculation. Nevertheless, base two or base e (or others) does not affect the calculation. In Equation (12), the numerator and the denominator will be different for different bases. However, we have proposed dimension as a ratio reflecting scale invariance. As long as the logarithm of the numerator and denominator has the same base, the result will not change. Therefore we can just use $log$ for ease of convenience.

\noindent \textbf {Example 1}: A framework of discernment is $\Theta=\left \{\omega_{1},\omega_{2}\right \}$, a mass function is 
 
 $$m(\omega_{1})=\frac{5}{6},\ m_{1}(\omega_{1},\omega_{2})=\frac{1}{6}$$.
According to Equation (9) and Equation (12),
 
 $$H_{D}=-\frac{5}{6}log(\frac{5}{6})-\frac{1}{6}log(\frac{\frac{1}{6}}{2^{2}-1})=0.9142$$.

 $$D_{m}=\frac{0.9142}{log[(2^{1}-1)^{\frac{5}{6}}+(2^{2}-1)^{\frac{1}{6}}]}=\frac{0.9142}{1.1381}=0.8033$$.

 \noindent \textbf {Example 2}: Given a framework of discernment $\Theta$, $(|\Theta|=1,2,...,20)$. A mass function is $m(\Theta)=1$. The result is show in Table 1. In Fig. 1, $x$ axis is Deng entropy and $y$ axis is $log(2^{|\Theta|}-1)^{1}$.

\begin{table}[!htbp]
\caption{The convergence process of Example 2 ($m(\Theta)=1$)}
\label{tab:2}       
\centering
\begin{tabular}{cccc}
\hline\noalign{\smallskip}
$|\Theta |$ & $H_{D}$ &  $log\sum_{i}(2^{|A_{i}|}-1)^{m(A_{i})}$ & $D_{m}$   \\
\noalign{\smallskip}\hline\noalign{\smallskip}
1 & 0 & 0 & 0 \\
2 & 1.5850 & 1.5850 & 1 \\
3 & 2.8074 & 2.8074 & 1 \\
4 & 3.9069 & 3.9069 & 1 \\
5 & 4.9542 & 4.9542 & 1\\
6 & 5.9773 & 5.9773 & 1\\
7 & 6.9887 & 6.9887 & 1\\
8 & 7.9944 & 7.9944 & 1 \\
...& ... & ... & ...\\
19 & 18.9999 & 18.9999 & 1 \\
20 & 20.0000 & 20.0000 & 1 \\

\noalign{\smallskip}\hline
\end{tabular}
\end{table}

\begin{figure}[!htbp]
\centering
  \includegraphics[width=3.5in]{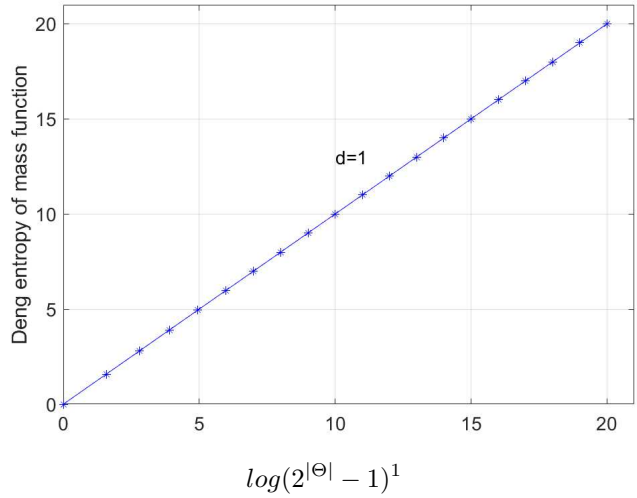}
$$log(2^{|\Theta|}-1)^{1}$$

\caption{The result of Example 2, where $|\Theta|=1,2,...,20$}
\label{fig:1}  
\end{figure}

As can been seen from Table 1, the dimension of $|\Theta|=1$ is 0. It means the complexity of a definite information is 0. Fig. 1 indicates a linear relationship between Deng entropy and the size of split of mass function when $|\Theta|=2,3,...,20$. The value of the slope is 1 and this means that the information dimension of the total uncertainty case is 1.

\begin{figure}[!htbp]
\centering
  \includegraphics[width=3.5in]{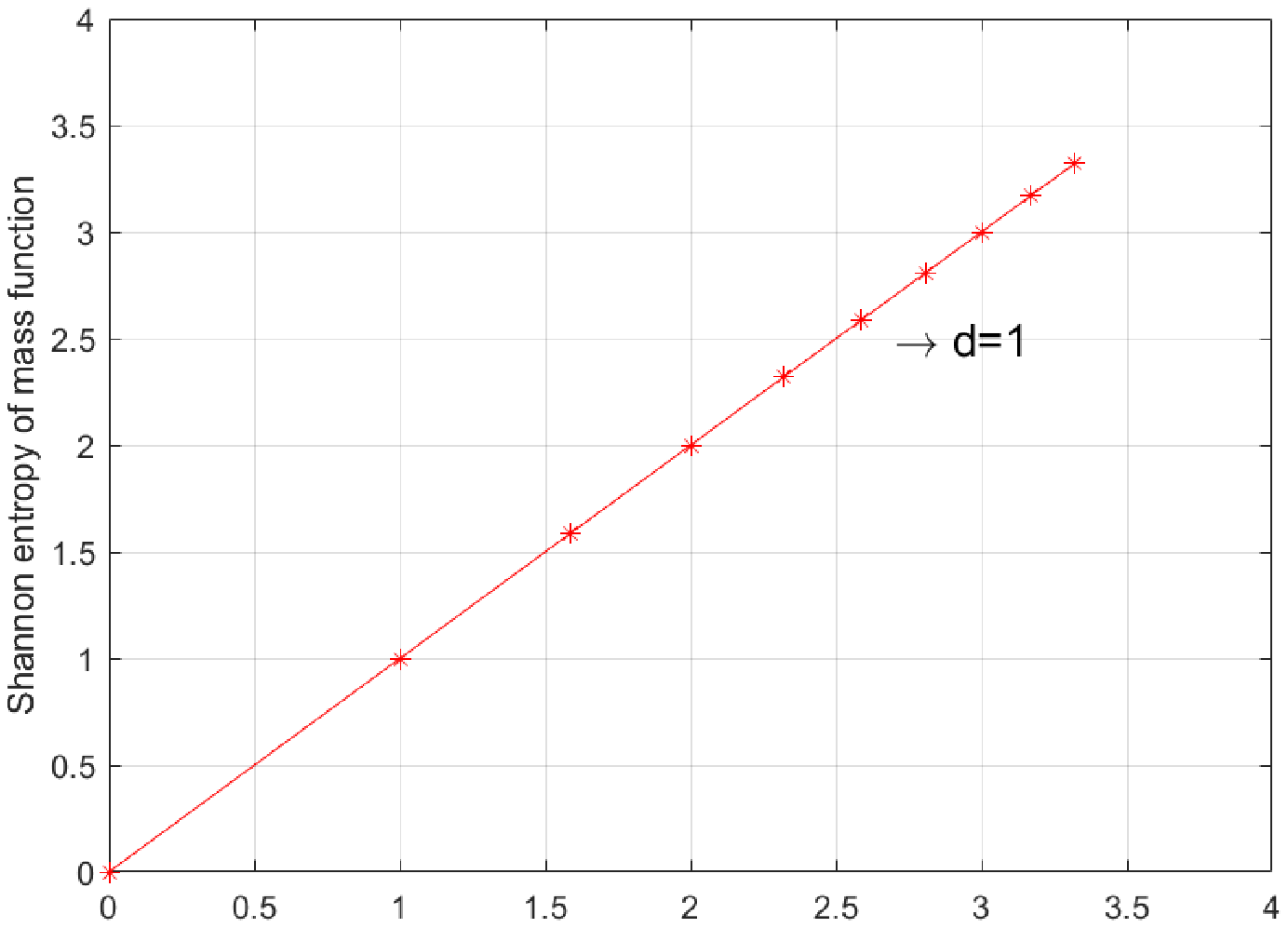}
$$log(N)$$

\caption{The result of Example 3, where $N=|\Theta|=1,2,...,10$}
\label{fig:2}  
\end{figure}

\noindent \textbf {Example 3}: Consider a framework of discernment $\Theta$, ($|\Theta|=1,2,...10$). A mass function is $(\left \{A_{1},A_{2},...A_{N}\right \},\ m(A_{i})=\frac{1}{N}))$, where $|A_{i}|=1$. The information dimension of this mass function is 1. The results of $D_{p}$ are listed in Table 2 and Fig. 2.
\begin{table}[!htbp]
\caption{The convergence process of Example 3 ($m(A)=\frac{1}{N}$)}
\label{tab:3}       
\centering
\begin{tabular}{cccc}
\hline\noalign{\smallskip}
$|\Theta |$ & $H_{S}$ &  $log\sum_{i}(2^{|A_{i}|}-1)^{P_{i}}$ & $D_{p}$   \\
\noalign{\smallskip}\hline\noalign{\smallskip}
1 & 0 & 0 & 0 \\
2 & 1 & 1 & 1 \\
3 & 1.5850 & 1.5850 & 1 \\
4 & 2 & 2 & 1 \\
5 & 2.3219 & 2.3219 & 1\\
6 & 2.5850 & 2.5850 & 1\\
7 & 2.8074 & 2.8074 & 1 \\
8 & 3 & 3 & 1 \\
9 & 3.1699 & 3.1699 & 1 \\
10 & 3.3219 & 3.3219 & 1\\

\noalign{\smallskip}\hline
\end{tabular}
\end{table}

Compared with Example 2 and Example 3, $m(\Theta)=1$ means total uncertainty in evidence theory and average distribution in probability theory $m(A_{i})=\frac{1}{N}$ has equal information dimension. According to Equation (18), the calculation of Example 3 is 
\begin{equation}
    D_{p}=\frac{\sum_{i}-p_{i}log(p_{i})}{log\sum_{i}(2^{|A_{i}|}-1)^{P_{i}}}=\frac{-\frac{1}{N}log(\frac{1}{N})\times N}{log(1\times N)}.
\end{equation}

The calculation of Example 2 according Equation (12) is
\begin{equation}
    D_{m}=\frac{H_{D}}{log\sum_{i}(2^{|A_{i}|}-1)^{m(A_{i})}}=\frac{-1\times log(\frac{1}{2^{N}-1})}{log(2^{N}-1)^{1}}.
\end{equation}

Equation (20) can be rewritten as
\begin{equation}
    D_{m}=\frac{-\frac{1}{2^{N}-1}log(\frac{1}{2^{N}-1})\times (2^N-1)}{log(1\times (2^N-1))}.
\end{equation}

From above Equation (19) and Equation (21), the case of $m(\Theta)=1$, $|\Theta|=N$ and the case of average distribution in probability, where the number of elementary events are $2^{N}-1$, are equivalent in expressing the complexity of information.

\noindent \textbf {Example 4}: Given a framework of discernment $\Theta$, $|\Theta|=N=1,2,...,25$. Its power set is $2^{\Theta}=\{\emptyset,,A_{1},\ A_{2},...,\ A_{2^{N}-1}\} $. A mass function with average assignment in power set is $m(A_{i})=\frac{1}{2^{N}-1}$. As can be seen from Table 3, with the increase of the size of $\Theta$, $D_{m}$ is changed but eventually goes to 1.5. Different from Example 2 and Example 3, which is a constant from the beginning, we assume that for this example the convergent value is the final information fractal dimension of mass function with average distribution in power set.

\begin{table}[!htbp]
\caption{The convergence process of Example 4 ($m(A_{i})=\frac{1}{2^{N}-1}$)}
\label{tab:4}       
\centering
\begin{tabular}{cccc}
\hline\noalign{\smallskip}
$|\Theta |$ & $H_{D}$ &  $log\sum_{i}(2^{|A_{i}|}-1)^{m(A_{i})}$ & $D_{m}$   \\
\noalign{\smallskip}\hline\noalign{\smallskip}
1 & 0 & 0 & 0 \\
2 & 2.1133 & 1.7834 & 1.1850 \\
3 & 3.8877 & 2.9691 & 1.3094 \\
4 & 5.5500 & 4.0186 & 1.3811 \\
5 & 7.1610 & 5.0260 & 1.4248\\
6 & 8.7428 & 6.0214 & 1.4520\\
7 & 10.3048 & 7.0418 & 1.4690\\
8 & 11.8523 & 8.0095 & 1.4798 \\
9 & 13.3886 & 9.0058 & 1.4867 \\
10 & 14.9162 & 10.0034 & 1.4911\\
...& ... & ... & ...\\
21 & 31.4965 & 21.0000 & 1.4998\\
22 & 32.9974 & 22.0000 & 1.4999 \\
23 & 34.4981 & 23.0000 & 1.4999 \\
24 & 35.9985 & 24.0000 & 1.4999 \\
25 & 37.4989 & 25.0000 & 1.5000\\

\noalign{\smallskip}\hline
\end{tabular}
\end{table}

\begin{figure}[!htbp]
\centering
  \includegraphics[width=3.5in]{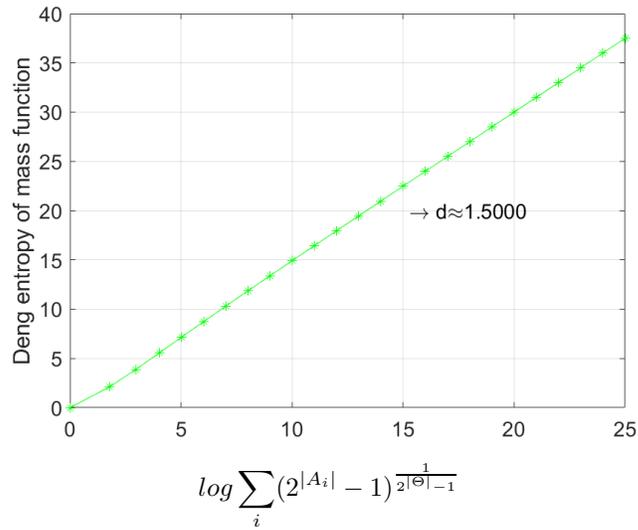}
$$log\sum_{i}(2^{|A_{i}|}-1)^{\frac{1}{2^{|\Theta|}-1}}$$
\caption{The result of Example 4, where $|\Theta|=1,2,...,25$}
\label{fig:4}  
\end{figure}

\begin{table}[!htbp]
\caption{The convergence process of Example 5 ($m(A)=\frac{2^{|A|}-1}{\sum({2^{|A_{i}|}-1})}$)}
\label{tab:5}       
\centering
\begin{tabular}{cccc}
\hline\noalign{\smallskip}
$|\Theta |$ & $H_{maxD}$ &  $log\sum_{i}(2^{|A_{i}|}-1)^{m(A_{i})}$ & $D_{m}$   \\
\noalign{\smallskip}\hline\noalign{\smallskip}
1 & 0 & 0 & 0 \\
2 & 2.3219 & 1.9757 & 1.1752 \\
3 & 4.2479 & 3.1071 & 1.3672 \\
4 & 6.0224 & 4.0970 & 1.4699 \\
5 & 7.7211 & 5.0679 & 1.5235\\
6 & 9.3772 & 6.0434 & 1.5516\\
7 & 11.0077 & 7.0265 & 1.5666 \\
8 & 12.6223 & 8.0157 & 1.5747 \\
9 & 14.2266 & 9.0091 & 1.5791 \\
10 & 15.8244 & 10.0052 & 1.5816\\
11 & 17.4178 & 11.0029 & 1.5830 \\
12 & 19.0084 & 12.0016 & 1.5838 \\
13 & 20.5971 & 13.0009 & 1.5843 \\
14 & 22.1845 & 14.0005 & 1.5846 \\
15 & 23.7711 & 15.0003& 1.5847\\
16 & 25.3572 & 16.0001 & 1.5848\\
17 & 26.9429 & 17.0001 & 1.5849 \\
18 & 28.5283 & 18.0000 & 1.5849 \\
19 & 30.1136 & 19.0000 & 1.5849 \\
20 & 31.6988 & 20.0000 & 1.5849\\

\noalign{\smallskip}\hline
\end{tabular}
\end{table}

\begin{figure}[!htbp]
\centering
  \includegraphics[width=3.5in]{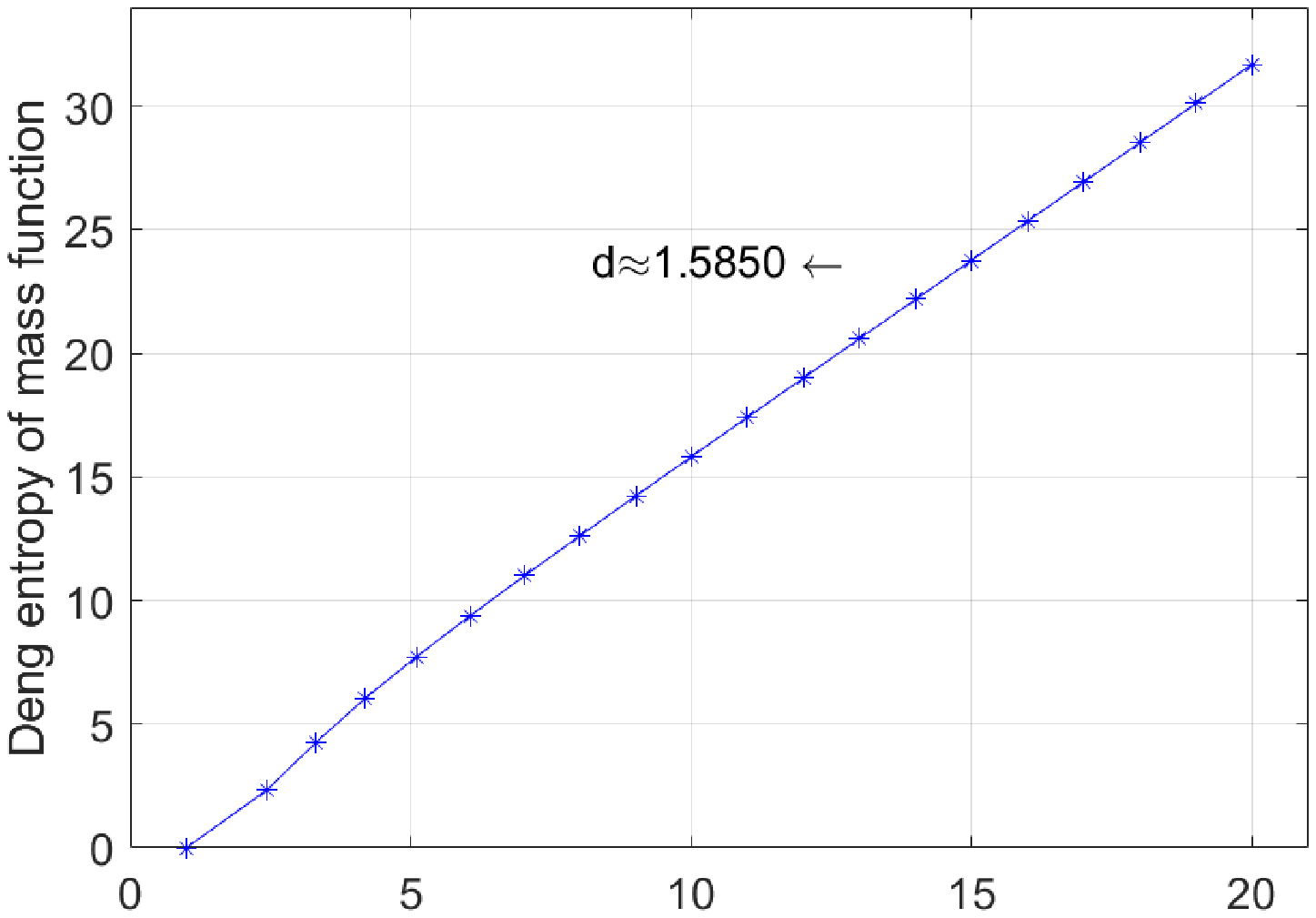}
$$log\sum_{i}(2^{|A_{i}|}-1)^{m(A_{i})}$$

\caption{The result of Example 5, where $|\Theta|=1,2,...,20$}
\label{fig:9}  
\end{figure}

\noindent \textbf {Example 5}: Given a framework of discernment $\Theta$, $|\Theta|=1,2,...,20$ and a mass function with maximum Deng entropy: $m(A)=\frac{2^{|A|}-1}{\sum({2^{|A_{i}|}-1})}$. Fig. 4 show the result and Fig. 5 is a Sierpinski triangle. 

From Table 4, with the size of $\Theta$ increasing, $D_{m}$ is a convergent sequence. The dimension of mass function with the maximum Deng entropy is $1.585$, which is the well-known fractal dimension of Sierpiński triangle.

\begin{figure}[!htbp]
\centering
  \includegraphics[width=3in]{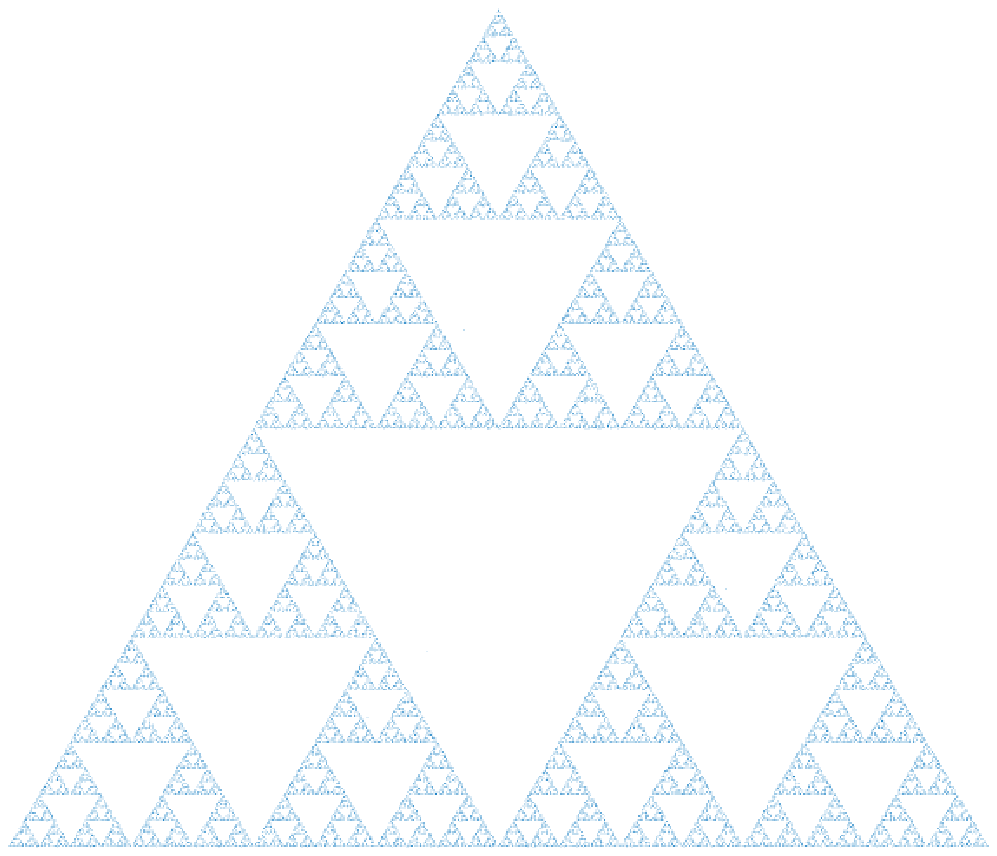}
\centering \caption{Sierpinski
triangle, whose fractal dimension is $\frac{ln3}{ln2}\approx 1.585$. Interested readers can refer to Ref.\ \cite{peitgen2006chaos} for the full construction steps.}
\label{fig:5}  
\end{figure}
\begin{figure}[!htbp]
\centering
  \includegraphics[width=3.5in]{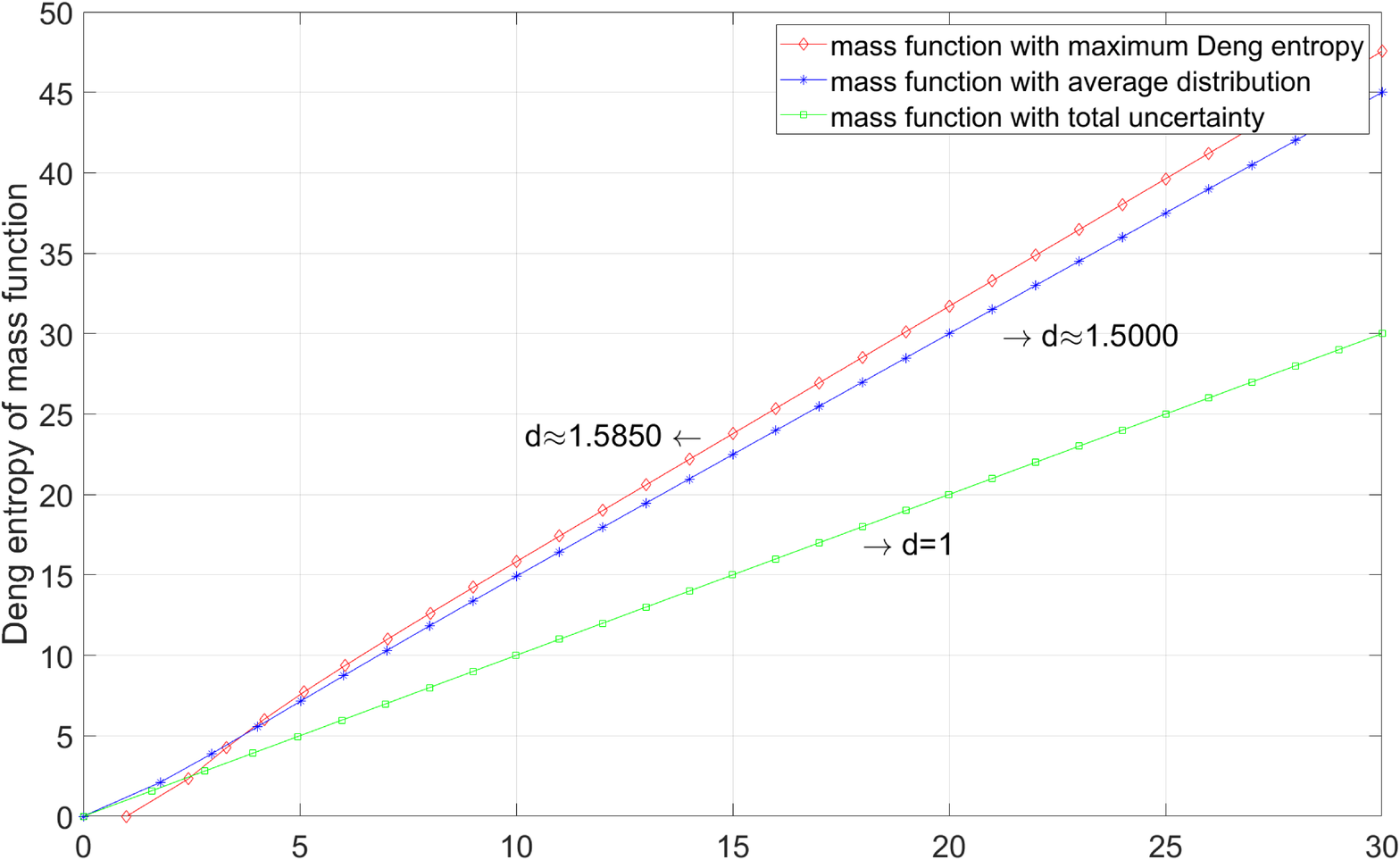}

\caption{Information dimension of Example 2, 4 and 5}
\label{fig:5}  
\end{figure}

Finally, we conclude the results from Example 2, Example 4 and Example 5 with a picture. Fig. 6 shows that there is a linear relationship between Deng entropy and the size of split of mass function. However, given any mass function like Example 1, is there a special distribution form of mass function that has the same dimension? What does the calculated dimension actually mean? 

There is still no common agreement on the interpretation of dimension, but one plausible explanation is related to the degree of freedom. In Euclidean Space, one-dimensional means that particle can only move in one direction. In two-dimensional space, particle can move in two orthogonal directions, and for three-dimensional, particle can move in three orthogonal directions. The higher the dimension, the more directions a particle can move; more variables are needed to measure it. However, we postulate here to express information fractal dimension as complexity. That is to say, the larger the information dimension is, the more complex the information represented by mass function will be. With regards to the property of fractal, the proposed information dimension can be applied to pattern recognition and multicriteria decision making in highly uncertain environment, in which collected information is incomplete and fragmentary. 

 \section{Conclusion}

How to determine the fractal dimension of uncertain information is still an open problem. In this paper, an information fractal dimension of mass function is proposed. The proposed method not only can calculate the dimension of probability distribution in probability theory, but also the mass function in evidence theory as well. Some interesting properties have been discussed. Importantly, we discover that the dimension of mass function with the maximum Deng entropy is 1.585, which is the same as the fractal dimension of Sierpinski triangle.

\section*{acknowledgements}

Yong Deng greatly appreciate China academician of the Academy of Engineering, Professor Shan Zhong and Professor You He, for their encouragement to do this research. Yong Deng greatly appreciates Professor Yugeng Xi to support this work. Ph.D student, Lipeng Pan, discussed the fractal dimension of Sierpiński triangle. This work has been continuously funded for the past ~20 years by grants such as the National Natural Science Foundation of China, Grant Nos. 30400067, 60874105, 61174022, 61573290 and 61973332, Program for New Century Excellent Talents in University, Grant No. NCET-08-0345, Shanghai Rising-Star Program Grant No.09QA1402900, Chongqing Natural Science Foundation for distinguished scientist, Grant No. CSCT, 2010BA2003 and JSPS Invitational Fellowships for Research in Japan (Short-term).

\bibliographystyle{ws-fnl}
\bibliography{mybibfile}

\end{document}